\documentclass[a4paper, amsfonts, amssymb, amsmath, reprint, showkeys, nofootinbib, twoside]{revtex4-1}
\usepackage[english]{babel}
\usepackage[utf8]{inputenc}
\usepackage[colorinlistoftodos, color=green!40, prependcaption]{todonotes}
\usepackage{amsthm}
\usepackage{mathtools}
\usepackage{physics}
\usepackage{xcolor}
\usepackage{graphicx}
\usepackage[left=23mm,right=13mm,top=35mm,columnsep=15pt]{geometry} 
\usepackage{adjustbox}
\usepackage{placeins}
\usepackage[T1]{fontenc}
\usepackage{lipsum}
\usepackage{csquotes}

\usepackage{amssymb}
\usepackage{natbib}

\usepackage[LGRgreek]{mathastext} 
\newcommand\tab[1][1cm]{\hspace*{#1}} 

\usepackage{lineno}
\usepackage{setspace}
\usepackage{hyperref}

\begin{document}
\title{Low temperature synthesis of BiFeO$_3$ nanoparticles with enhanced magnetization and promising photocatalytic performance  in dye degradation and hydrogen evolution }

\author{M. A. Basith\textit{$^{*}$}, Nilufar Yesmin, and Rana Hossain}
    \affiliation{Nanotechnology Research Laboratory, Department of Physics, Bangladesh University of Engineering and Technology, Dhaka-1000, Bangladesh.\\
    \textsuperscript{*}Corresponding author: mabasith@phy.buet.ac.bd\\
    \\ DOI: \href{https://pubs.rsc.org/en/content/articlehtml/2018/ra/c8ra04599b}{10.1039/c8ra04599b}}


\begin{abstract}
In this investigation, we have synthesized BiFeO$_3$ nanoparticles by varying hydrothermal reaction temperatures from 200 $^\circ$C to 120 $^\circ$C to assess their visible-light driven photocatalytic activity along with their applicability for hydrogen production via water splitting. The rhombohedral perovskite structure of BiFeO$_3$ is formed for hydrothermal reaction temperature up to 160 $^\circ$C, however, for a further decrement of reaction temperature a mixed sillenite phase is observed. The XRD Rietveld analysis, XPS analysis and FESEM imaging ensure the formation of single-phase and well crystalline nanoparticles at 160 $^\circ$C reaction temperature with 20 nm of average size. The nanoparticles fabricated at this particular reaction temperature also exhibit improved magnetization, reduced leakage current density and excellent ferroelectric behavior. These nanoparticles demonstrate considerably high absorbance in the visible range with low band gap (2.1 eV). The experimentally observed band gap is in excellent agreement to the calculated band gap using the first-principles calculations. The favorable photocatalytic performance of these nanoparticles has been able to generate more than two times of solar hydrogen compared to that produced by bulk BiFeO$_3$ as well as commercially available Degussa P25 titania. Notably, the experimentally observed band gap is almost equal for both bulk material and  nanoparticles prepared at different reaction temperatures. Therefore, in solar energy applications, the superiority of BFO nanoparticles prepared at 160 $^\circ$C reaction temperature may be attributed not only to solely their band gap but also to other factors, such as reduced particle size, excellent morphology, well crystallinity, large surface to volume ratio, ferroelectricity and so on.
\end{abstract}


\maketitle

\section{Introduction}
Nanotechnology has made tremendous impact in contribution to the solution of local as well as global energy crisis by converting sunlight directly into electric/chemical power. A promising area for the application of nanotechnology is the hydrogen economy.  As a non-polluting source of energy, hydrogen (H$_2$) can be a good alternative fuel of the future \cite{ref1,ref2,ref3} without polluting the environment. H$_2$ production can be performed from different domestic feedstock including hydrocarbon fossil fuels  and water. Notably, H$_2$ produced by splitting of water \cite{ref11001, ref11002, ref11003} under solar light illumination are of great promise as a fuel that is essentially carbon free and inexhaustible in nature with potential for generating low cost power in power plants as well as in vehicles of next generation.

\tab Photocatalytic disintegration of water \cite{ref7003} into H$_2$ and O$_2$ while irradiated by light in the visible region has been considered  as one of the promising routes for H$_2$ production as an endless source of clean fuel for many applications. In 1972, Fujishima and Honda used successfully titania (TiO$_2$) photoelectrode for the decomposition of water \cite{ref3} under illumination with sunlight and without any applied electric power. Since then many propositions along with numerous researches have been conducted on hydrogen production via water splitting \cite{ref4}. In the past decade, extensive studies were performed on a broad range of materials, including modified TiO$_2$ \cite{ref5}, SrTiO$_3$ \cite{ref6} and metal oxide-metal hybrids like ZnO/(La,Sr)CoO$_3$ \cite{ref7} and CdS/TiO$_2$/Pt \cite{ref8} to improve their photocatalytic water splitting ability under UV-visible light irradiation. Recently, significant amount of research is being done with a view to improving the photocatalytic activity of the multiferroic materials such as bismuth iron oxide (BiFeO$_3$) to investigate its potentiality for solar hydrogen generation \cite{ref9,ref1}. BiFeO$_3$ (BFO) intrigues with its multiferroic properties being potentially applicable in energy-related problems especially for photocatalytic hydrogen production imputed to its relatively small band-gap (2.6 eV) \cite{ref11,ref12,ref13, ref14}. It is reported that BiFeO$_3$ generates hydrogen at a greater amount than  commercially available titania under exact parametric conditions \cite{ref9}.  Utilization of a somewhat broader sunlight spectrum along with  a large polarization value of BFO efficiently generated electron-hole pairs \cite{ref101}. Notably, BFO could potentially supplant the widely investigated photocatalyst TiO$_2$ that has an inherent limitation of wide band gap (3.2 eV), qualifying its utilization to only 4\% of the solar spectrum \cite{ref102}. Despite the great potential it possesses, the somewhat inferior photocatalytic performance of bulk BiFeO$_3$ restrains its commercialization \cite{ref15, ref16}. 

\tab Multiferroic BiFeO$_3$ was discovered in 1960, however, it remains difficult to synthesize single phase BFO nanoparticles with excellent morphology owing to the fact that the temperature range for stability is in fact quite narrow for this material. If the synthesis condition, particularly the processing temperature is not subject to accurate control, other impurity phases such as Bi$_2$Fe$_4$O$_9$ appears in BFO nanoparticles. Therefore, a number of synthesis techniques for BFO nanoparticles have been developed at different temperatures. These include soft chemical routes (e.g. sol-gel method) \cite{ref23}, ultrasonication technique \cite{ref17}, microemulsion technique \cite{ref24} and microwave-hydrothermal process \cite{ref25}. Most of these synthesis techniques, inherently cost inefficient, comprise of higher than 400 $^o$C calcination temperatures for achieving phase purity for BFO which introduces irregular morphology and broad particle size distribution. While sol-gel has gained popularity among different wet chemical techniques, it has a stringent requirement of an annealing temperature of 600 $^o$C as the concluding step as well as a post treatment similar to solid-state synthesis techniques for elimination of impurity phases. Combustion synthesis can yield nanosized BFO powder at the expense of phase purity \cite{ref27} being inferior to the ones synthesized by the soft chemical route \cite{ref23}. Hence, synthesis of phase pure BFO nanoparticles at a moderate preparation condition has become quite challenging. Since many years, hydrothermal synthesis of phase pure BFO powders gained intensive research interest due to its potential in synthesizing crystalline ceramics at a temperature of 200 $^\circ$C  or even lower without any further step of calcination. This is a cost effective and simple method, and the primacy of this technique is its significantly lower required temperature than both solid-state and sol-gel syntheses. During the preparation of BFO, such a low processing temperature qualifies the reactants from getting volatilized as well as minimizing the calcination introduced impurities.

\tab Therefore, it is obvious that enormous effort was paid to conduct extensive research to synthesize pure BFO nanoparticles using hydrothermal technique. However, there is still lack of a thorough investigation on synthesis, characterization and application of pure BFO nanoparticles with desired phase and crystallinity prepared at very low temperature using a facile, low cost synthesis route like hydrothermal. It is indeed very essential to shed light to know different crystallographic phases formed at different reaction temperatures and finally figure out the appropriate reaction temperature to obtain phase pure BFO nanoparticles ensuring their quality as required for practical applications. In this investigation, we delineate the synthesis of BFO nanoparticles adapting a facile hydrothermal process for different reaction temperatures and a fixed KOH concentration of 10 M \cite{ref75}. With a view to knowing the crystallographic phases, we scrutinized Rietveld refinement on powder XRD data of the samples synthesized over a wide range of temperatures varying from 200 $^\circ$C to 120 $^\circ$C. Notably, in most of the previous hydrothermal investigations either crystallographic phases of BFO nanoparticles were not identified by Rietveld or related analysis and/or it was done at a particular temperature \cite{ref75, ref200} not like here over a wide range of temperatures.  We observed that well-dispersed, single-phase perovskite BFO nanoparticles with average particle size 20 nm can be produced by hydrothermal process at temperature as low as  160 $^\circ$C (from here this material is referred as BFO 160 unless otherwise specified). A further reduction of temperature to 140 $^\circ$C  did not allow fabrication of single phase perovskite structure rather produce sillinite type mixed phases nanoparticles. Interestingly, the single-phase BFO 160 nanoparticles  demonstrate excellent multiferroic, optical and photocatalytic properties. The hydrogen evolution experiment exhibited that BFO 160 nanoparticles generated more than two times hydrogen than that of BFO bulk material and commercially available TiO$_2$ under the same experimental conditions. 

\section{Experimental details}
\subsection{Sample Preparation}
In the present investigation, the chemical reagents are analytical grade bismuth nitrate (Bi(NO$_3$)$_3$.5H$_2$O), ferric nitrate (Fe(NO$_3$)$_3$.9H$_2$O), and potassium hydroxide (KOH). In a typical procedure, 1mmol bismuth nitrate pentahydrate [Bi(NO$_3$)$_3$.5H$_2$O] and 1mmol ferric nitrate nonahydrate [Fe(NO$_3$)$_3$.9H$_2$O] in stoichiometric proportions were dissolved in 50mL of 10M KOH solution and stirred it for 4h \cite{ref75, ref26, ref200}. The mixture was transferred to a Teflon-lined autoclave (stainless steel, capacity 100 ml) and was heated at different reaction temperatures between 200 $^\circ$C to 120 $^\circ$C for 6 hours followed by natural cooling to bring it to room temperature. Centrifugation was applied to collect the final products and consequent rinsing was performed with distilled water and ethanol. Prior to further characterization, they were dried at 90 $^\circ$C for 6 hours. Bulk BiFeO$_3$ sample had also been prepared from high purity oxides using solid state reaction method with an objective of comparing its properties to the synthesized nanoparticles \cite{ref76}.

\subsection{Materials Characterization}
The crystal structures and phase composition of the synthesized samples were determined from XRD data using a diffractometer (PANalytical Empyrean) with a Cu X-ray source (Wavelength, ${\lambda}$: K$_{\alpha1}$= 1.540598 \AA  and K$_{\alpha2}$ = 1.544426 \AA. The particle size along with morphology was observed with help of field emission scanning electron microscopy (FESEM) (XL30SFEG; Philips, Netherlands and S4300; HITACHI, Japan). X-ray photoelectron spectroscopy (XPS) was used to investigate the chemical states of synthesized nanoparticles. A vibrating sample magnetometer (VSM) was used for measuring the magnetic properties of the BFO powders at room temperature. The leakage current density along with polarization vs electric field hysteresis loops were measured using a ferroelectric loop tracer in conjunction with external amplifier (10 kV). Pellets were prepared by pressing the powders with a hydraulic press followed by annealing at 750 $^\circ$C with high heating rate (20 $^\circ$C/min) with a view to measuring the electrical properties \cite{ref201}. An Ultraviolet-visible (UV-vis) spectrophotometer (UV-2600, Shimadzu) was used to obtain diffuse reflectance spectra (DRS) of the samples for wavelength ranging from 200 to 800 nm. BaSO$_4$ powder, a nonabsorbing standard material, was used as the reference for total reflectance with the focused beam spot size being approximately of 2 mm$^2$.
\subsection{Theoretical Calculation of Optical Properties}
With a view to understanding the optical properties of BFO bulk and BFO 160 nanoparticles, the optical absorption was calculated based on crystallographic structural parameters obtained from Rietveld refinement \cite{ref701} of this investigation. The calculations were performed using first-principles density functional theory (DFT) within the plane wave pseudopotential (PWPP) framework as implemented in the Cambridge Serial Total Energy Package (CASTEP) \cite{ref702}. The generalized gradient approximation (GGA) with the Perdew-Burke-Ernzerhof (PBE) exchange-correlation functional within ultrasoft pseudopotentials (USP) was implemented to describe the electron-ion interaction of valance electrons of  \( Bi \left( 6s^{2}6p^{3} \right)  \) ,  \( Fe \left( 3d^{6}4s^{2} \right)  \)  and  \( O \left( 1s^{2}2p^{4} \right)  \) \cite{ref703}. Plane wave cutoff and k-point sampling, tested with a cutoff of 400 eV and a k-point sampling of 6$ \times $6$ \times $6, was found sufficient for the rhombohedral unit cell of the samples. Spin polarized mode during self-consistent field (SCF) calculations was endorsed and a SCF tolerance of  \( 10^{-7} \) eV/atoms was used. To assure a precise agreement with experimental results, the on-site Coulomb interaction was included in the DFT+ U approach with U = 3 eV to 7 eV with interval 0.5 eV for Fe 3d orbital \cite{ref704}. The optical absorption coefficient was obtained by the equation \( ~ \alpha =\sqrt[]{2 \mu _{0} \omega ~}\sqrt[]{~\sqrt[]{ \varepsilon _{1}^{2} \left(  \omega  \right) + \varepsilon _{2}^{2} \left(  \omega  \right) - \varepsilon _{1} \left(  \omega  \right) }} \), where  \(  \varepsilon _{1} \left(  \omega  \right)  \)  and  \(  \varepsilon _{2} \left(  \omega  \right)  \)  are frequency dependent real and imaginary parts of dielectric function,  \(  \omega  \)  is photon frequency, \( ~ \mu _{0} \)  is the permeability of free space \cite{ref705}. The real part of the dielectric function  \(  \varepsilon _{1} \left(  \omega  \right)  \)  can be evaluated from the imaginary part  \(  \varepsilon _{2} \left(  \omega  \right)  \)  by the famous Kramers-Kronig relationship \cite{ref706}. Direct optical band gap of the samples was obtained from the equation \( ~ \left(  \alpha h \upsilon  \right) ^{n}=A \left( h \upsilon -E_{opt} \right) \), where A is a constant and n denotes the transition type as follows: n = 2 for direct allowed, 2/3 for direct forbidden, 1/2 for indirect allowed and 1/3 for indirect forbidden transitions \cite{ref707}. By extrapolating the linear proportion of the  \(  \left(  \alpha h \upsilon  \right) ^{n}~vs~h \upsilon  \)  plot to  \(  \left(  \alpha h \upsilon  \right) ^{n}=0 \) the optical band gap of the samples was calculated using Tauc plot \cite{ref708}.

\subsection{Photocatalytic performance}
The photocatalytic performance of BFO bulk material and hydrothermally prepared nanoparticles at different reaction temperatures were subject to evaluation by photodegradation of Rhodamine B (RhB) \cite{ref7004} under illumination of visible light in aqueous solutions using a 500 W Xe lamp as a solar simulator accompanied with a visible cutoff filter (${\lambda}$ $\geq$ 420 nm). Beginning with a concentration of 15 mg/L of RhB, 80 mg of catalyst was later added  to the solution in a quartz glass reactor for each experiment. Before illumination, the mixture was magnetically stirred in the dark for 1 hour for achieving the equilibrium between adsorption and desorption processes. After sufficient agitation, 4 mL of mixture was taken followed by centrifugation at 5000 rpm for 10 min so that the catalyst powders are removed. The concentration of RhB was then estimated from the measurement of the maximum absorbance at 553 nm with a UV-visible spectrophotometer. Testing for stability, the remaining photocatalyst powders in suspension were separated by centrifugation after photocatalytic degradation of RhB and distilled water was used to wash them to remove the residual RhB. Prior to using them in another photocatalytic reaction, they were further dried. The same procedure was followed three times. For comparison commercially available Degussa P25 titania nanoparticles were used to perform dye degradation experiment under the same experimental conditions. 

\subsection{Hydrogen Generation}
A photocatalytic hydrogen generation experiment has been performed in a slurry-type photochemical reactor. In a typical experiment, 20mg of catalyst particles is measured and initiated in the reaction container with 30ml water. Magnetic stirring is applied to mix the solution while purging the system with argon gas for 30min to ensure an atmosphere, necessarily inert in nature as required for the splitting process. A 500W Xenon lamp was used for photoillumination. The gas was collected when each of the experiments ended and scrutinized  in the gas chromatography (GC) device that is equipped with a thermal conductivity detector (TCD) and a gas analyzer for identifying the gas components. The GC programming, set up with a reverse polarization, made the hydrogen peaks occur in an upwards direction so that we can make a comparative analysis of the peak intensities of the different gases that were produced. The hydrogen evolution test was also performed  for commercially available Degussa P25 titania nanoparticles under the same experimental conditions.

\section{Results and discussions}
\subsection{Structural characterization}
\begin{figure*}
 \centering
 \includegraphics[width= 1\textwidth]{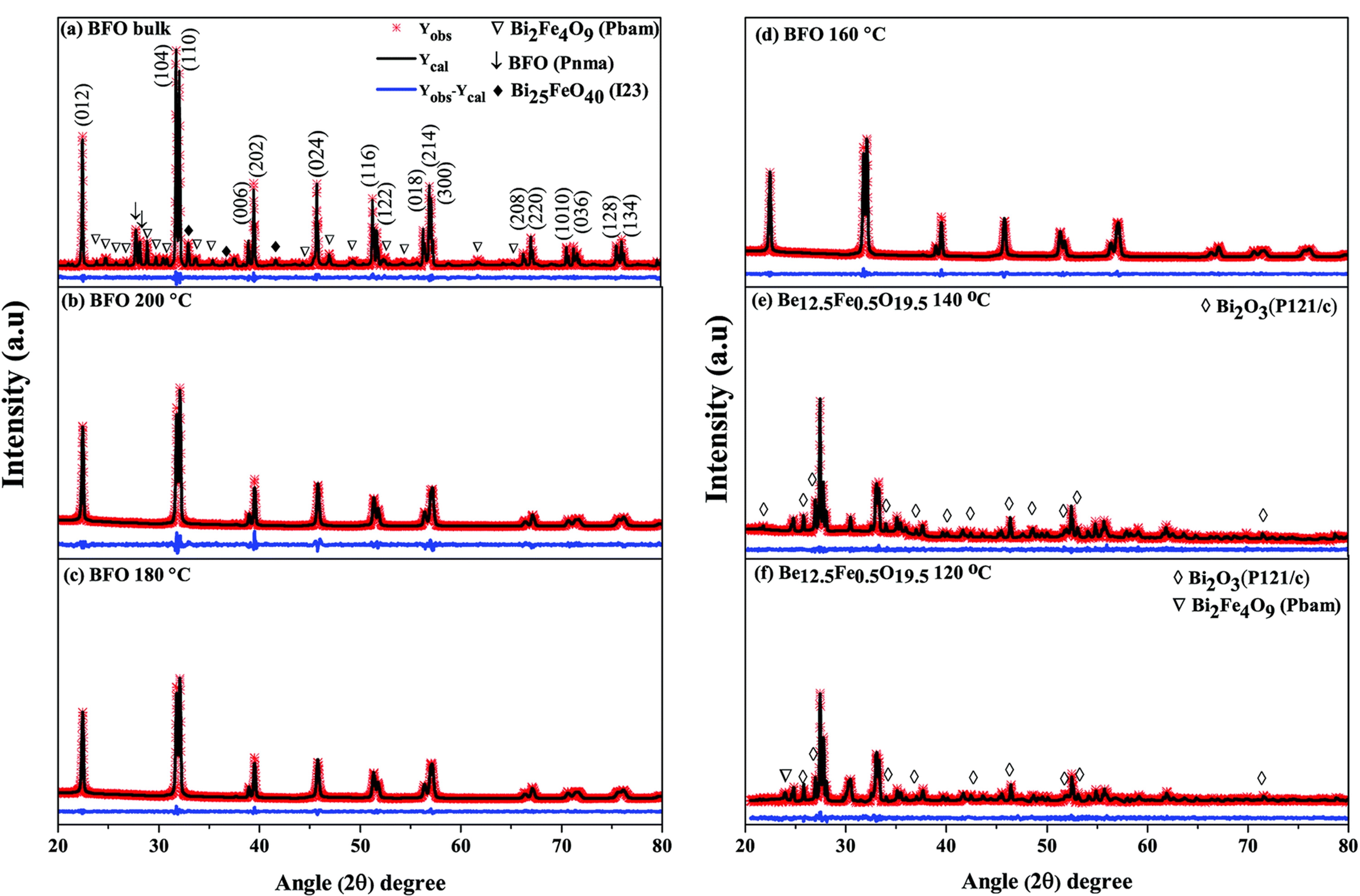}
 \caption{Rietveld plots of XRD patterns of (a) BFO bulk material synthesized by solid state reaction technique and hydrothermally prepared nanoparticles at reaction temperatures (b) 200 $^\circ$C (c) 180 $^\circ$C (d) 160 $^\circ$C (e) 140 $^\circ$C (f) 120 $^\circ$C.} \label{fig:fig1}
\end{figure*}

Fig. \ref{fig:fig1} shows the Rietveld refined XRD patterns of BFO nanoparticles synthesized by hydrothermal process at reaction temperatures 200 $^\circ$C, 180 $^\circ$C, 160 $^\circ$C, 140 $^\circ$C and 120 $^\circ$C with the peaks indexed on the basis of Rhombohedral structure (JCPDS File No. 71-2494). For the comparison, XRD pattern of BFO bulk polycrystalline sample  prepared via conventional solid state reaction is also depicted in Fig. \ref{fig:fig1} (a). Rietveld refinement was performed with the FULLPROF package \cite{ref301}. The structural variables and constituent phases (in wt\%) calculated from the Rietveld refined XRD spectra of the synthesized materials have been provided in Supplemental Table 1 \cite{ref401}.  The small values of Rwp and Rp  shown in Supplemental Table 1 \cite{ref401} indicate very good fit for hydrothermally prepared BFO nanoparticles to the Rhombohedral structure. Reproduction of all the observed reflections can be performed from this structural model. Calculated lattice parameters (a and c) have been listed in Table 1 for all the major phases.

\tab As was mentioned, the major phase of the synthesized materials is observed to be of Rhombohedral crystal structure belonging to space group R3c. However, some secondary impurity phases are present for both bulk polycrystalline material and hydrothermally prepared nanoparticles at reaction temperatures 140 $^\circ$C and 120 $^\circ$C. For nanoparticles prepared at hydrothermal reaction temperatures 200 $^\circ$C to 160 $^\circ$C no obvious impurity phase (like Fe$_2$O$_3$, Fe$_3$O$_4$, Bi$_2$Fe$_4$O$_9$ etc,) has been determined to the precision limit of XRD. Moreover, it is seen that for hydrothermal reaction temperature up to 160 $^\circ$C, the nanoparticles (Fig. \ref{fig1} (b-d)) present a single-phase structure.  For a further decrease of reaction temperature to 140 $^\circ$C and 120 $^\circ$C a structural transition to sillenite type Bi$_{25}$FeO$_{40}$ (Fig. \ref{fig:fig1} (e-f)) phase is formed. Thus, the structural scrutiny  clearly indicates that the minimal reaction temperature for obtaining perovskite type pure BFO phase is 160 $^\circ$C. Therefore, we have carried out further characterization for nanoparticles prepared hydrothermally at reaction temperatures of 200 $^\circ$C-160 $^\circ$C. 

\tab From the refinements of the XRD patterns, the Fe-O-Fe bond angle and FeO$_6$ octahedral tilt angle $\omega$ are recapitulated in Table 1. The Fe-O-Fe bond angle of BFO bulk material and nanoparticles prepared  at 200 $^\circ$C, 180 $^\circ$C and 160 $^\circ$C reaction temperatures are 160.8753$^\circ$, 158.6353$^\circ$, 158.3842$^\circ$, 152.8377$^\circ$, respectively as shown in Table 1. The average Fe-o-Fe bond angle is maximum for bulk BFO material, however, a monotonic decrease of the average Fe-O-Fe angle from 158.6353$^\circ$ to 152.8377$^\circ$ is observed with decrease of hydrothermal reaction temperatures. Correspondingly, we observed an increase of the FeO$_6$ octahedral tilt angle ($\omega$) with the decrease of reaction temperatures. The alterations in the Fe-O-Fe bond angle and FeO$_6$ octahedral tilt angle play role to modify the magnetic behavior of the synthesized nanoparticles.

\begin{table*}
	\small
	\caption[centering]{Structural parameters determined from XRD spectra}
	\resizebox{1\textwidth}{!}{
	\begin{tabular}{|l|l|l|l|l|l|}
		\hline
		Sample & a=b & c & Volume &bond angle&FeO$_6$ tilt \\  
		& (\AA) & (\AA)& (\AA)$^3$& Fe-O-Fe ($^\circ$)&angle ($\omega$ in $^\circ$)\\   \hline
		BFO Bulk&5.5775(1)&13.8663(1)& 373.577&160.8753&10.310\\ \hline
		BFO (200 $^\circ$C)&5.5743(1)&13.8612(3)& 373.003&158.6353&10.682\\ \hline	
		BFO (180 $^\circ$C)&5.5761(1)&13.8634(3)&  373.314&158.3842&10.807\\ \hline
		BFO (160 $^\circ$C)&5.5758(1)&13.8656(4)& 373.327&152.8377&15.857\\ \hline
	\end{tabular}}\\
\end{table*}

\subsection{Morphology analysis}
\begin{figure}
 \centering
 \includegraphics[width= 0.5\textwidth]{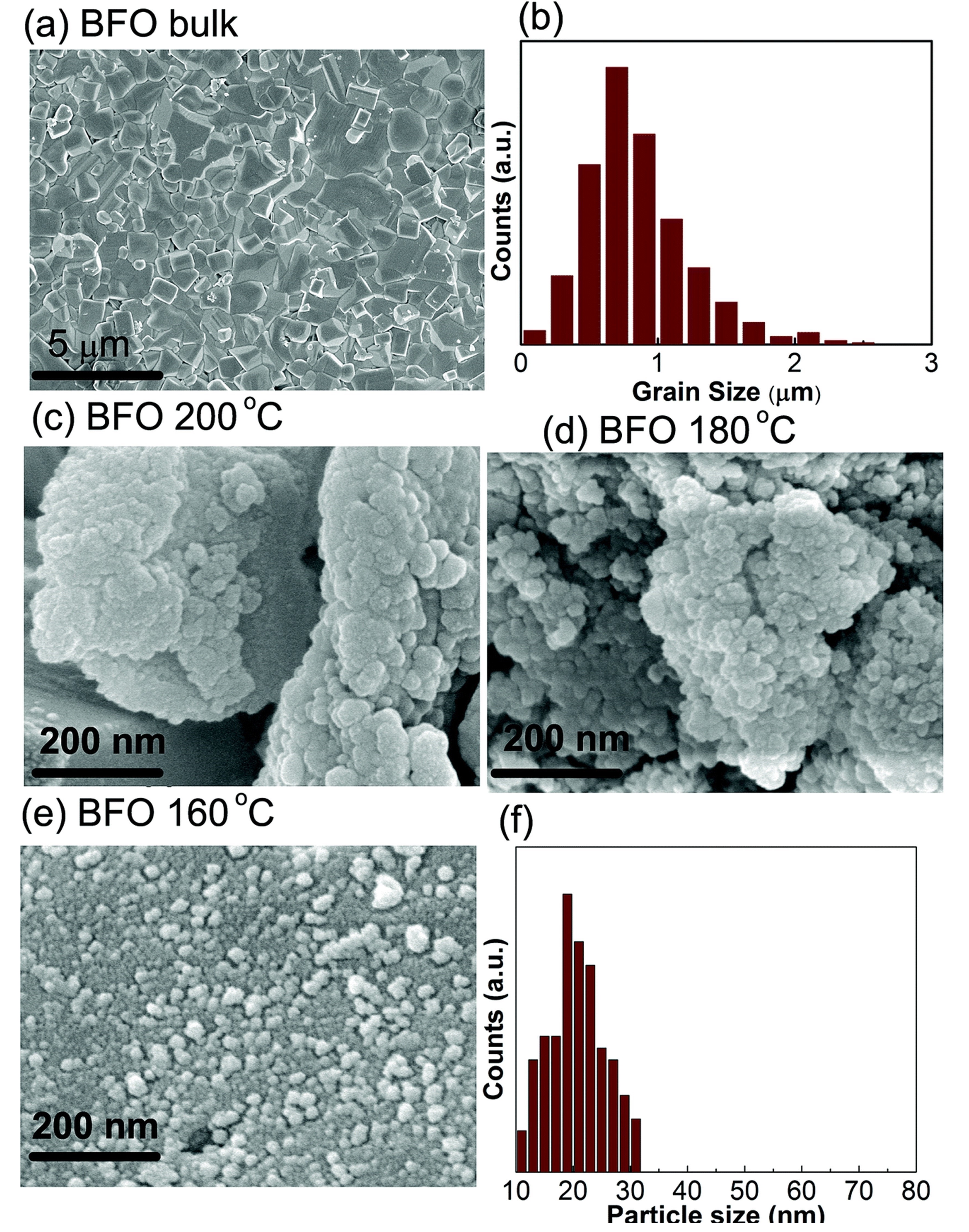}
 \caption{FESEM images of surface morphology of (a) BFO bulk material synthesized by solid state reaction technique and (b) corresponding histogram. Hydrothermally prepared nanoparticles at reaction temperatures (c) 200 $^\circ$C (d) 180 $^\circ$C, respectively. The nanoparticles and their corresponding histogram prepared at hydrothermal reaction temperature 160 $^\circ$C are depicted in (e) and (f), respectively.}
 \label{fig:fig2}
\end{figure}

The morphology of the surface and the size of the fabricated particles were analyzed by using FESEM imaging. The surface morphology and their corresponding histograms of bulk BFO sample produced by solid state reaction technique and BFO materials synthesized by hydrothermal method at 200 $^\circ$C, 180 $^\circ$C and 160 $^\circ$C reaction temperatures are shown in Fig. \ref{fig:fig2}. According to Fig. \ref{fig:fig2} (a) and corresponding histogram Fig. \ref{fig:fig2} (b), the average grain size of bulk BFO material is around 1 $\mu$m. The size of the bulk BiFeO$_3$ was also determined from the XRD pattern (maximum intensity peak) using Scherrer equation and found to be around 600 nm which is comparable to the value obtained from FESEM image (Fig. \ref{fig:fig2} (a) and (b)). FESEM images in figure \ref{fig:fig2} (c-d) show BFO materials synthesized for the reaction temperatures 200 $^\circ$C, and 180 $^\circ$C, respectively. Fig. \ref{fig:fig2} (e-f) depicts the BFO nanoparticles synthesized at 160 $^\circ$C reaction temperature and their corresponding histogram for size distribution.

\tab The electronic micrographs obtained by FESEM, Fig. \ref{fig:fig2} (c-d), show the particles are prone to get interconnected with one another and aggregating to larger particles at higher reaction temperatures i.e. at 200 $^\circ$C and 180 $^\circ$C. However, for a comparatively lower reaction temperature of 160 $^\circ$C, spherical and well dispersed BFO nanoparticles, Fig. \ref{fig:fig2} (e-f) with sizes ranging from 20 nm to 30 nm is formed. Our investigation demonstrates that  the morphology of the synthesized BiFeO$_3$ materials changed from agglomerations to regular spherical shape with a narrow distribution of size when the reaction temperature is reduced from 200 $^\circ$C to 160 $^\circ$C. Previous investigation also reported such a change in morphology of BFO nanoparticles by varying the concentration of KOH for a fixed hydrothermal reaction temperature of 200 $^\circ$C \cite{ref881}. We think, from energy consumption point of view it is much more desirable to control the surface morphology at a lower hydrothermal reaction temperature of  160 $^\circ$C under a fixed concentration of KOH. Moreover, Bi is highly volatile and such a low processing temperature prevents reactants from getting volatilized during the formation of BFO powders and minimize the amount of impurities as was mentioned earlier.

During the hydrothermal preparation of BFO powders, the size of the final products depends on two mutually competitive processes: crystal nucleation and crystal growth \cite{ref882}. Provided the rate of crystal nucleation is greater than that of growth, then the product particle size becomes smaller. We expect that in our experiment with the decrease of reaction temperature to 160 $^\circ$C, the growth rate was decreased, however, nucleation rate was increased and consequently the average size of the particles was significantly reduced. The smaller particle size may provide an increased surface area to volume ratio and substantial adsorption capacity, that is required to obtain a high photocatalytic performance.

\subsection{X-ray photoelectron spectroscopic analysis}
\begin{figure*}
 \centering
 \includegraphics[width = 1\textwidth]{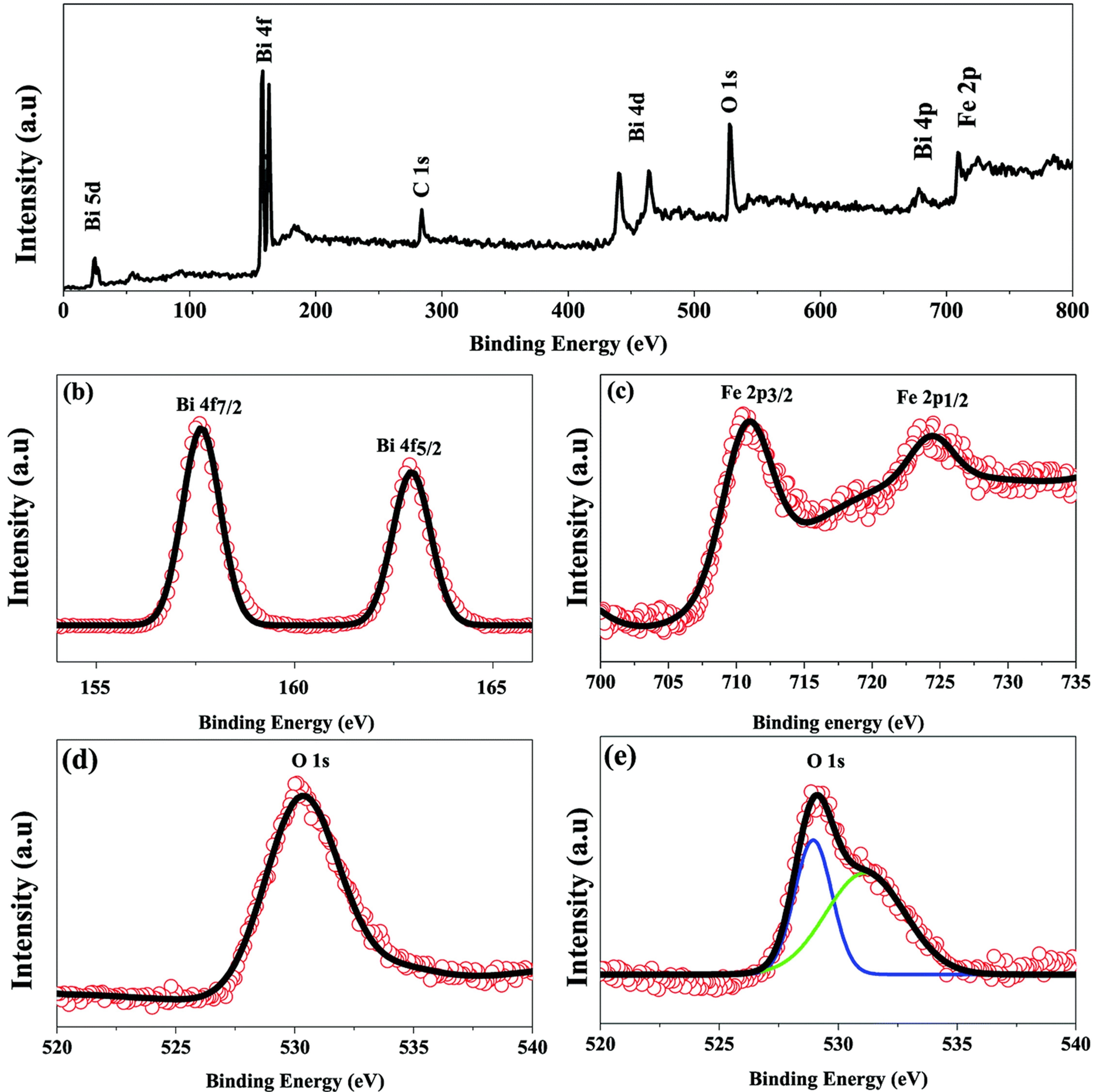}
 \caption{XPS spectra of the as-synthesized BFO 160 nanopowders (a) a typical XPS survey spectrum, (b–d) high-resolution XPS core spectra of Bi 4f, Fe 2p, and O 1s, respectively. (e) XPS spectra of O 1s of BFO bulk materials.}
 \label{fig:fig3}
\end{figure*}

The XRD analysis and FESEM imaging confirmed that single phase, well crystalline BFO nanoparticles formed at 160 $^\circ$C reaction temperature may meet the demand of practical applications. Therefore, X-ray photoelectron spectroscopy was carried out to further identify the chemical composition of BFO 160 nanoparticles which were prepared at hydrothermal reaction temperature 160 $^\circ$C. Fig. \ref{fig:fig3} shows the typical full survey of XPS spectrum of BFO 160 nanoparticles. Observing the survey spectrum, aside from the weak peak of C 1s, only O, Fe, and Bi core levels were detected, indicating that the resulting product is highly pure. Fig. \ref{fig:fig3} (b-d) shows the high resolution XPS  spectra of the Bi 4f, Fe 2p and O 1s core levels for the BFO 160 nanoparticles, respectively. The core level XPS spectrum of Bi can be distinguished by two Gaussian peaks at 4 f$_{7/2}$ and 4 f$_{5/2}$ corresponding to the binding energy of 157.64 eV and 162.93 eV, respectively. The spin-orbit splitting energy of the Bi 4f is 5.3 eV \cite{ref883, ref884}. Fig. \ref{fig:fig3} (c) shows a high resolution of XPS measurements for Fe contents. Two main peaks: one at 710.9 eV and another at 724.7 eV have been assigned to corresponding states of Fe 2p$_{3/2}$ and 2p$_{1/2}$ with the energy for spin-orbit splitting being 13.35 eV for Fe 2p \cite{ref883}. For a better insight of the chemical composition we have scrutinized the O 1s peak of BFO 160 nanoparticles (Fig. \ref{fig:fig3} (d)) and bulk BiFeO$_3$ materials Fig. \ref{fig:fig3}(e). The uniformly single XPS peak of O 1s (530.29 eV) demonstrated that the BiFeO$_3$ nanoparticles synthesized at 160 $^\circ$C  hydrothermal reaction temperature prepared in this investigation is indeed a compound of single-phase. The XPS spectrum of O 1s core level for bulk BiFeO$_3$ materials, as shown in Fig.3 (e), can be de-convoluted into two symmetric Gaussian peaks. The lower binding energy peak at 529.8 eV corresponds to the O 1s core spectrum, while the higher binding energy peak is related to the oxygen vacancy in the bulk BFO sample.

\subsection{Magnetic characterization}
\begin{figure}
 \centering
 \includegraphics[width= 0.5\textwidth]{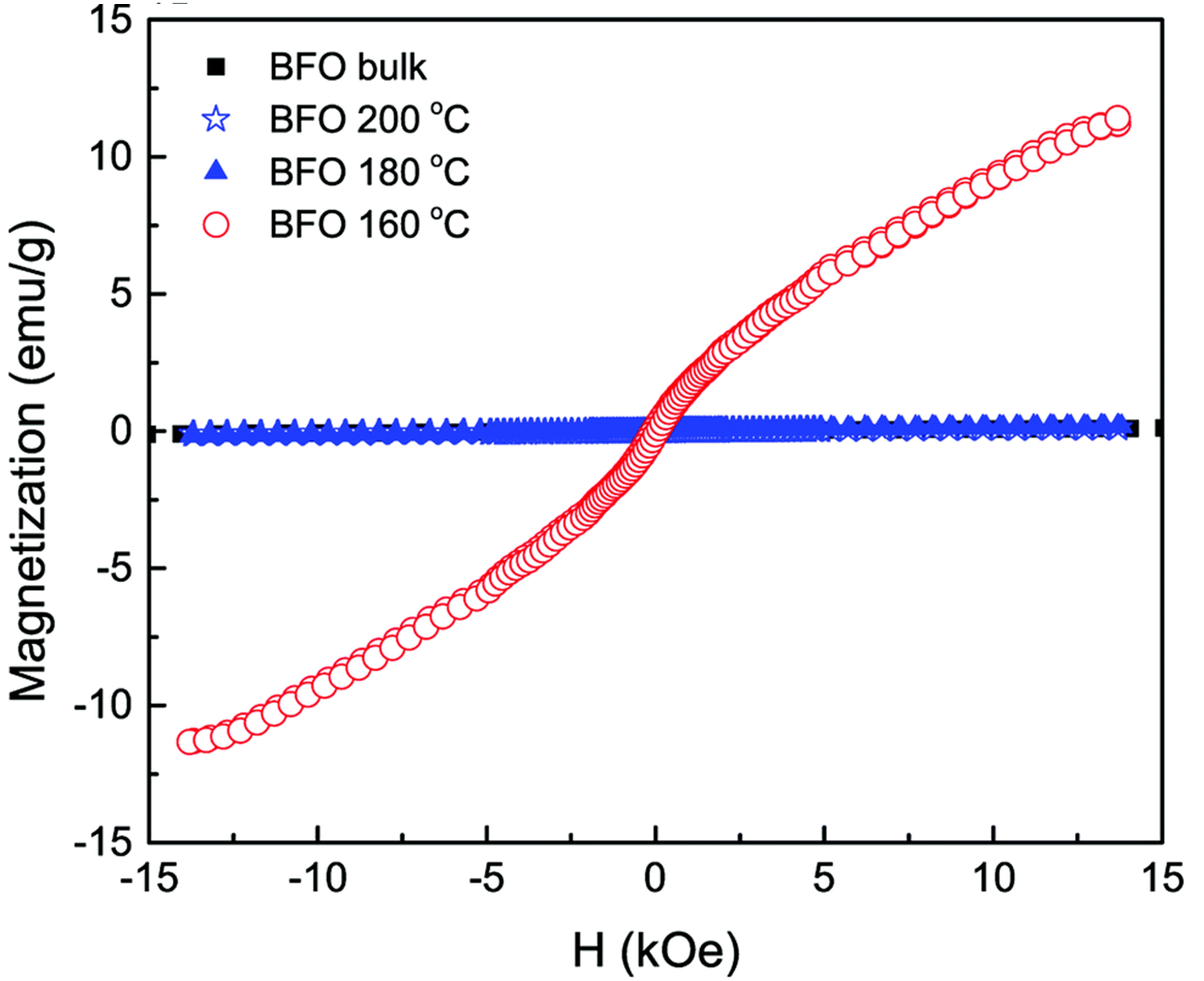}
 \caption{The room temperature M–H hysteresis loops of BFO bulk material synthesized by solid state reaction technique and hydrothermally prepared nanoparticles at reaction temperatures 200 $^\circ$C, 180 $^\circ$C and 160 $^\circ$C.}
 \label{fig:fig4}
\end{figure}

The field dependent magnetic hysteresis loops (M-H) of bulk BFO material and hydrothermally synthesized nanoparticles are investigated at room temperature with an applied magnetic field of up to $\pm$15 kOe. In Fig. \ref{fig:fig4}, the obtained magnetization for bulk BFO prepared by solid state reaction technique changes linearly with magnetic field which demonstrates its antiferromagnetic nature \cite{ref33,ref11113}. However, the magnetic behavior of BFO nanoparticles varies depending on hydrothermal reaction temperatures. We have calculated the remanent magnetization (M$_r$) of the synthesized samples from the M-H hysteresis loops \cite{ref500, ref11112}. The maximum magnetization (M$_s$)  at an applied magnetic field of 13 kOe is also calculated from the hysteresis loops. Both M$_r$ and M$_s$ are higher for BFO nanoparticles prepared at 160 $^\circ$C hydrothermal reaction temperature. In particular at 13 kOe applied magnetic field, the M$_s$ value is 11.3 emu/g which is exceptionally high compared to the values reported in literature \cite{ref501, ref502}. In investigation \cite{ref501}, BFO nanoparticles with a particle size of 14 nm were prepared using sol-gel method for annealing temperature of 400 $^\circ$C and the obtained M$_s$ was 1.55 emu/g at 50 kOe. The observed M$_s$ was 1.4 emu/g at 70 kOe for 18 nm BFO nanoparticles fabricated by sol-gel method with annealing temperature 425 $^\circ$C  \cite{ref502}. In our investigation, such a remarkable high value of magnetization (11.3 emu/g) of the hydrothermally synthesized nanoparticles at reaction temperature 160 $^\circ$C demands an extensive investigation of the magnetic behavior of this material. 

\tab The rationale behind improved magnetization can be imputed to several factors, such as: (i) The Rietveld refinement of XRD data (Table I) clearly reveals that the Fe-O-Fe bond angle is minimum for nanoparticles with average size 20 nm produced at hydrothermal reaction temperature 160 $^\circ$C. Such a decrement in Fe-O-Fe bond angle is crucial for developing ferromagnetism due to the strengthening of magnetic interaction caused by decreased Fe-O-Fe bond angle. This is consistent with previous investigations which reported that ferromagnetism can be realized in BFO by tuning Fe-O-Fe bong angle with reduced particle size \cite{ref503, ref504}. (ii) At room temperature,  magnetization in G-type antiferromagnetic bulk BFO is subdued along with a spin helical ordering structure showing a periodicity of 62 nm. Compared to bulk materials, in the case of phase pure nanoparticles, the periodic spiral-regulated spin structure can be modified with the decrease in size of particles, and hence magnetization increases.(iii) The enhanced ferromagnetism in BFO nanoparticles with average particle 20 nm may be due to the reduced size of the phase pure nanoparticles. With decrease in particle size, the surface-to-volume ratio is enhanced and hence the surface spins contribute more to the arrant magnetic moment of the particle \cite{ref17, ref500, ref501, ref502}. 

\begin{figure*}
 \centering
 \includegraphics[width= 1\textwidth]{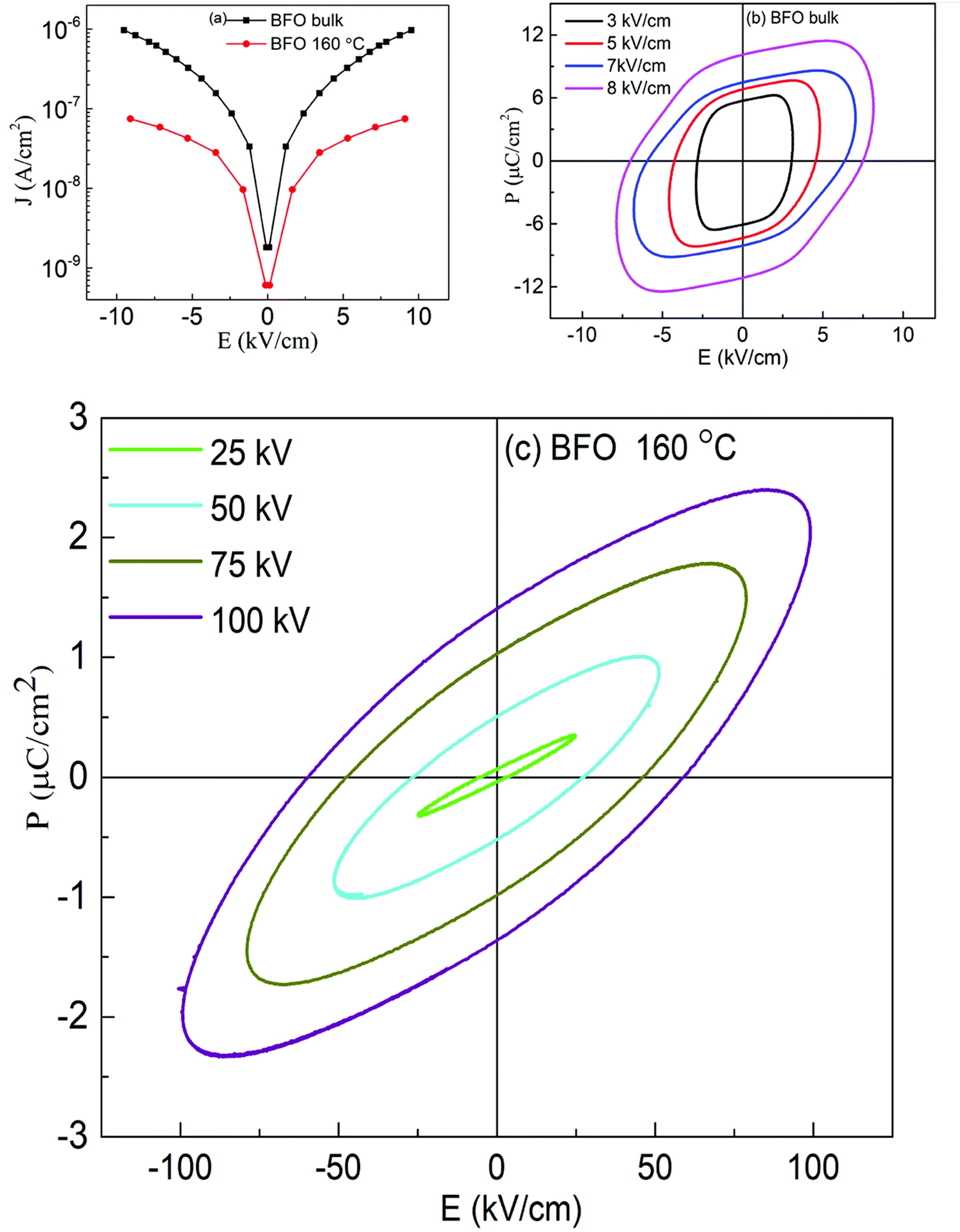}
 \caption{(a) Leakage current density of BFO bulk material and nanoparticles prepared at 160 $^\circ$C hydrothermal reaction temperature. The polarization vs. electric field (P–E) hysteresis loops of (b) BFO bulk material and (c) nanoparticles prepared at 160 $^\circ$C hydrothermal reaction temperature.}
 \label{fig:fig5}
\end{figure*}

\tab To further confirm the multiferroicity of hydrothermally prepared BFO 160 nanoparticles, P--E hysteresis loops were investigated. Prior to conducting P-E hysteresis, the leakage current density (J) versus electric field (E) measurements were carried out.

\tab From Fig. \ref{fig:fig5} (a), it can be implied that the leakage current density of the hydrothermally synthesized BFO 160 nanoparticles is smaller compared to that of the bulk. Impurity phases along with oxygen vacancies contribute to the high leakage current of bulk material \cite{ref601, ref502}. Previous demonstration from Fig. \ref{fig:fig1} (d) for BFO 160 nanoparticles indicate inhibition of impurity phases. Furthermore, Fig. \ref{fig:fig3} shows that in BFO 160 nanoparticles, we observed only O 1s core spectrum,  and obviously no oxygen vacancy related defects. Therefore, we think that the oxygen vacancies, principally induced by volatilized Bi$^{3+}$ ion in the bulk material, were suppressed in BFO 160 nanoparticles and consequently the density of leakage current was reduced.

\tab Fig. \ref{fig:fig5} (b-c) shows the  polarization of ferroelectric hysteresis loops of (b) bulk BFO and (c) BFO 160 nanoparticles measured for varying electric fields. Driving frequency maintained at 50 Hz, continuously increasing electric field resulted in the increase of remanent polarization as stronger electric field with its higher level of driving power contributed to the reversal of ferroelectric domains \cite{ref603}.

\tab Freely movable charges are expected to contribute more to the electrical hysteresis loop for bulk BFO. Thus, the bulk BFO material shows a round shaped P-E loop as observed in Fig. \ref{fig:fig5} (b) owing to high leakage current which is evident from Fig. \ref{fig:fig5} (a). For BFO 160 nanoparticles, the contribution of the dynamic charges to the polarization is reduced due to the decrease in rounded shapes of the loops. Furthermore, the P-E loops of BFO nanoparticles tend to become increasingly typical which can be associated with the reduction in leakage current density \cite{ref604} and decreased space charge defect \cite{ref605} compared to that of bulk material. Thus, the measurement of electrical properties shows an improvement in ferroelectric behavior of BFO 160 nanoparticles. While comparing the P-E loops between BFO bulk and BFO 160 nanoparticles, Fig. \ref{fig:fig5} (b) and (c) also demonstrate that the significantly improved breakdown voltage up to 100 kV/cm for BFO 160 nanoparticles compared to 10 kV/cm for bulk material.

\subsection{Optical characterization}
\begin{figure}
 \centering
 \includegraphics[width= 0.5\textwidth]{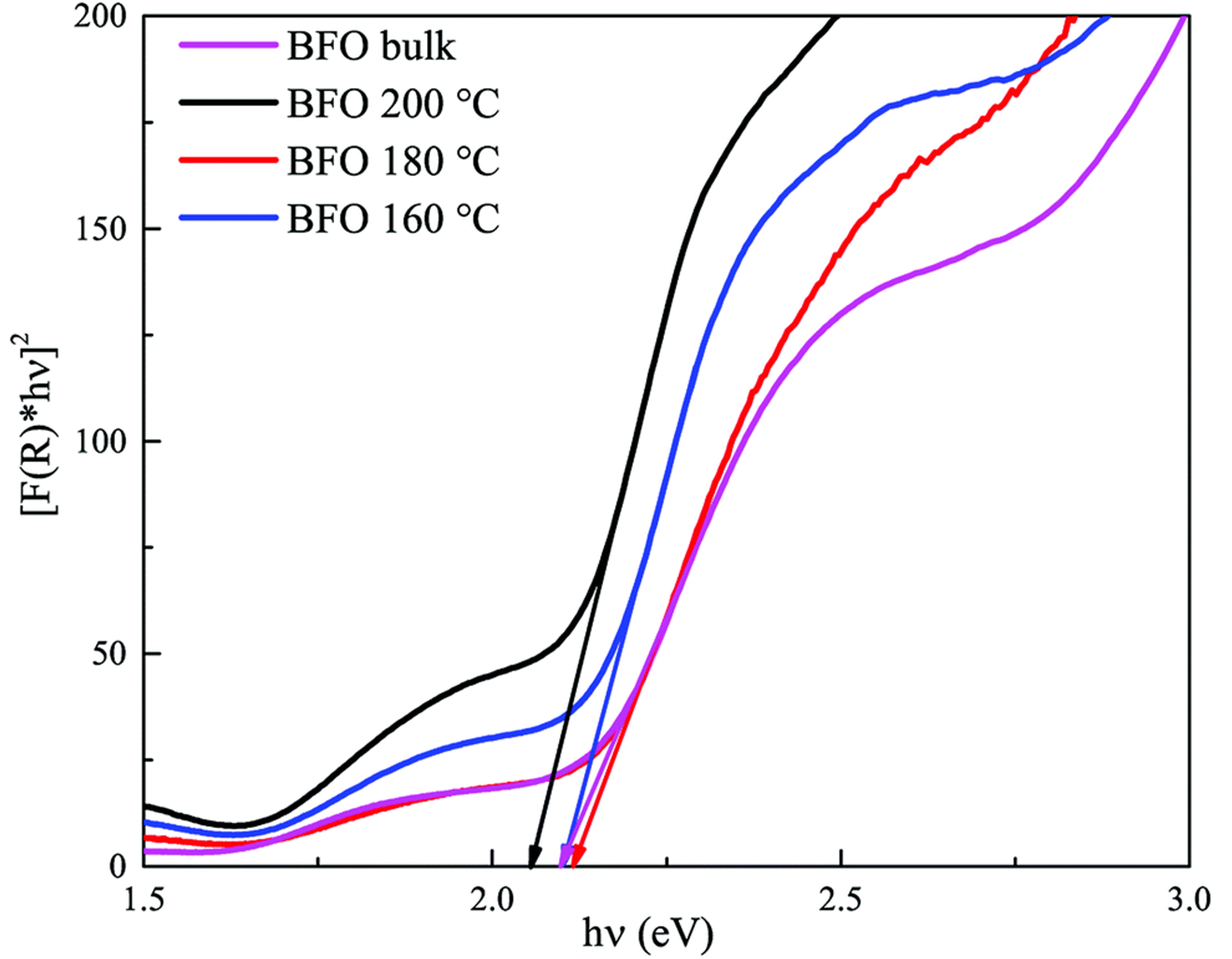}
 \caption{Tauc plots for band gap estimation of BFO bulk material synthesized by solid state reaction technique and hydrothermally prepared nanoparticles at reaction temperatures 200 $^\circ$C, 180 $^\circ$C and 160 $^\circ$C.}
 \label{fig:fig6}
\end{figure}

Photocatalytic performance of a material depends on its ability to effectively absorb visible light. Optical band gap signifies the minimum energy of the photons that a material can absorb to generate electron-hole pairs via interband transition. To determine the optical band gap, the diffuse reflectance spectra of the synthesized materials were obtained from UV-vis spectrophotometric measurements. Determining F(R), a parameter proportional to the absorption coefficient, from the diffuse reflectance value using Kubelka-Munk function, we can have an estimate of the optical band gap\cite{ref606},

\begin{equation}
F(R)=\frac{(1-R)^2} {2R}.
\end{equation}
The band gap of the materials can be calculated by forming the Tauc plot using the following equation \cite{ref606},
\begin{equation}
F(R)^{*}h\nu=A(h\nu-E_{g})^n,
\end{equation}

where, h$\nu$, A, and E$_g$ denote the energy of photons, constant of proportionality, and optical band gap, respectively. n is a number that takes the value of 1/2 and 2 for direct and indirect transitions, respectively. Since, BFO is widely regarded as a material with direct band gap, we have considered n=2 and obtained the Tauc plots accordingly as shown in Fig. \ref{fig:fig6}.

\tab Fig. \ref{fig:fig6} displays $[F(R)*h\nu]^{2}$ vs $h\nu$ curves for optical band gap calculation of BFO bulk materials and nanoparticles hydrothermally synthesized at different reaction temperatures. Extrapolating the linear region of the curve to the energy axis, we can determine the band gap. Optical band gap for bulk BFO material is found to be 2.1 eV. For hydrothermally prepared nanoparticles at reactions temperatures 200 $^\circ$C, 180 $^\circ$C and 160 $^\circ$C, the direct band gaps are 2.05 eV, 2.11 eV and 2.1 eV, respectively. This energy band gap is consistent with value reported in Ref. \cite{ref15,ref11217}. However, in those investigations, nanoparticles were produced at considerably higher annealing temperatures. The small band gap exhibited by synthesized BFO bulk and nanoparticles indicates their potential as good photocatalysts since they are capable of absorbing visible range photons effectively. However, all the synthesized materials have almost equal band gaps and hence, superiority of a particular sample in photocatalytic applications cannot attributed solely to its band gap. Rather, there are other factors which will also play important roles in determining the efficiency of a material as photocatalyst such as particle size, surface morphology, phase purity, ability to suppress electron-hole recombination etc.

\subsubsection{Theoretical calculation of band gap}
\begin{figure*}
 \centering
 \includegraphics[width= 1\textwidth]{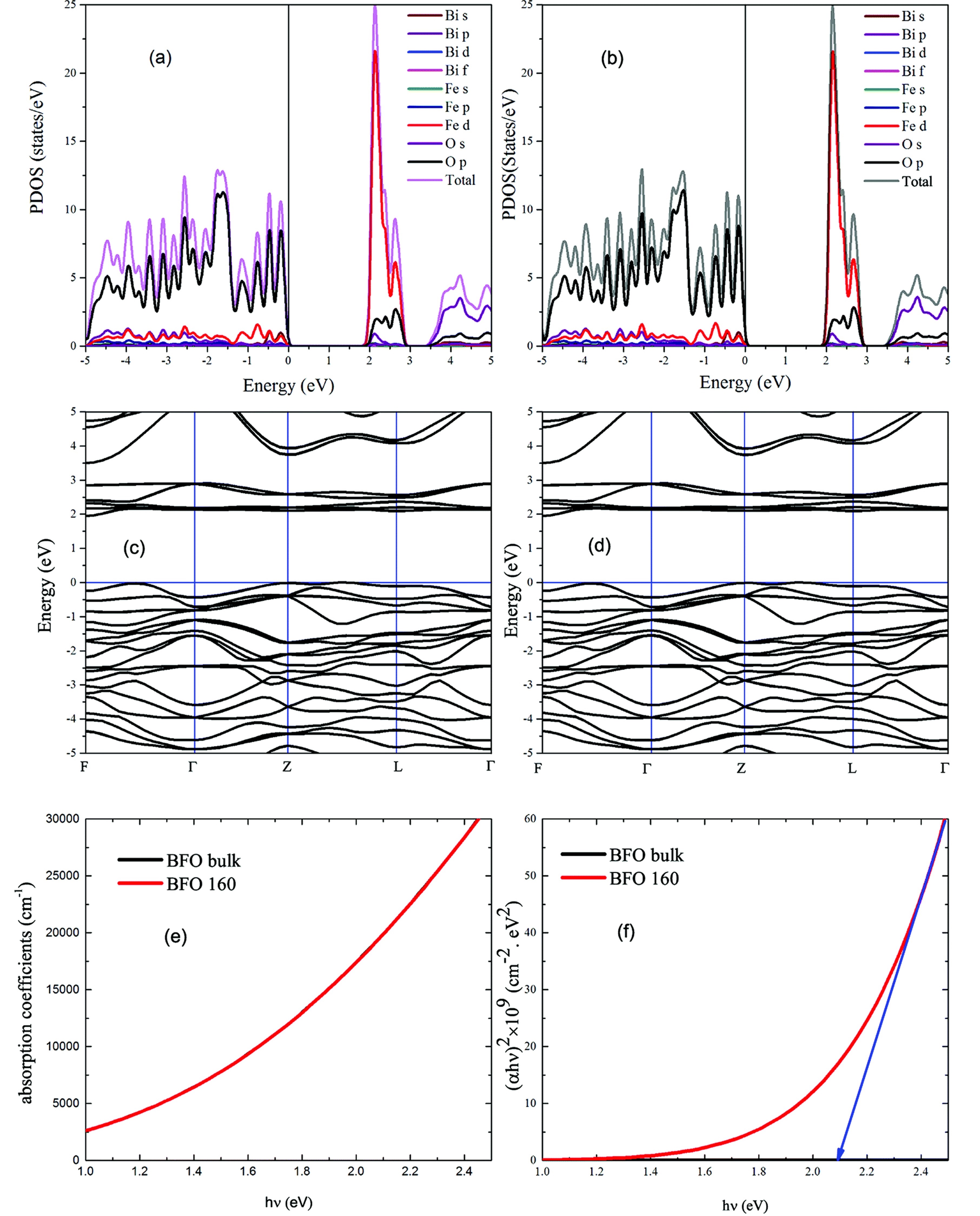}
 \caption{DFT calculated PDOS spectra of (a) BFO bulk and (b) BFO 160; band structures of (c) BFO bulk and (d) BFO 160; DFT calculated absorption spectra of (e) BFO bulk and BFO 160; (f) plot of DFT calculated ($\alpha$$h\nu$)\textsuperscript{2} vs. $h\nu$ for BFO bulk and BFO 160.}
 
\label{fig:fig7} 
\end{figure*}

BFO bulk and BFO 160 nanoparticles were subject to First-principles calculation with a view to comparing the experimentally obtained values of band gap to the theoretically calculated ones. While standard DFT methods are widely in use for such calculations, experimentally obtained outcomes of BiFeO$_3$ are more closely comparable to the DFT calculated values when Hubbard U parameter, a measure of effective on-site Coulomb interaction, is chosen carefully \cite{ref2003,ref709,ref711,ref712,ref713}. Experimentally obtained crystallographic parameters were refined by Rietveld method and their geometry was optimized prior to using them for DFT calculations. PDOS, electronic band structure and Optical absorption spectra of both BFO bulk and BFO 160 nanoparticles were obtained via DFT calculations using GGA-PBE functionals. While performing DFT calculations, U value was varied from 3.0 eV to 7.0 eV at 0.5 eV interval. DFT calculated absorption spectra closely match the experimentally obtained ones for U = 4.5 eV and therefore, further observations were carried out for this value of U. As we can observe from the PDOS spectra (Fig. \ref{fig:fig7}(a) and \ref{fig:fig7}(b)) of both BFO bulk and BFO 160, conduction band minima(CBM) is dominated by 3d orbital of Fe atoms and 2p orbital of oxygen atoms dominate the valence band maxima(VBM). We can infer from this observation that the electron transitions from VBM to CBM can be attributed to these two orbitals and there are no major d-d transitions involved. From the electronic band structures as shown in Fig. \ref{fig:fig7}(c) and \ref{fig:fig7}(d), VBM lies between Z and L point for BFO bulk and BFO 160 nanoparticles and hence the calculated direct band gap for both materials found to be 2.1 eV. Since, transition from 2p orbital of oxygen atoms to 3d orbitals of Fe atoms is allowed and direct, our use of n = 2 in the formation of Tauc plot from experimentally obtained absorption coefficients is justified. Fig. \ref{fig:fig7} (e) shows the theoretical optical absorption coefficients of BFO bulk and BFO 160 obtained from our DFT calculation. These absorption spectra were transformed into Tauc plots as shown in Fig.  \ref{fig:fig7} (f) and tangents were drawn to the linear region of the plot. Extrapolating the tangent to the energy axis, optical band gap of the corresponding material was found from the energy axis intercept. As observed from Fig. \ref{fig:fig7} (f), both BFO bulk and BFO 160 exhibit the same direct optical band gap of 2.1 eV. These theoretically calculated values are in excellent agreement with the ones experimentally determined (2.1 eV for both BFO bulk and BFO 160). With this agreement between theoretical and experimental calculation, we can conclude that there has been no significant change in the optical band gap for BFO bulk and nanoparticles.

\subsection{Photocatalytic activity}

\begin{figure*}
 \centering
 \includegraphics[width= 1\textwidth]{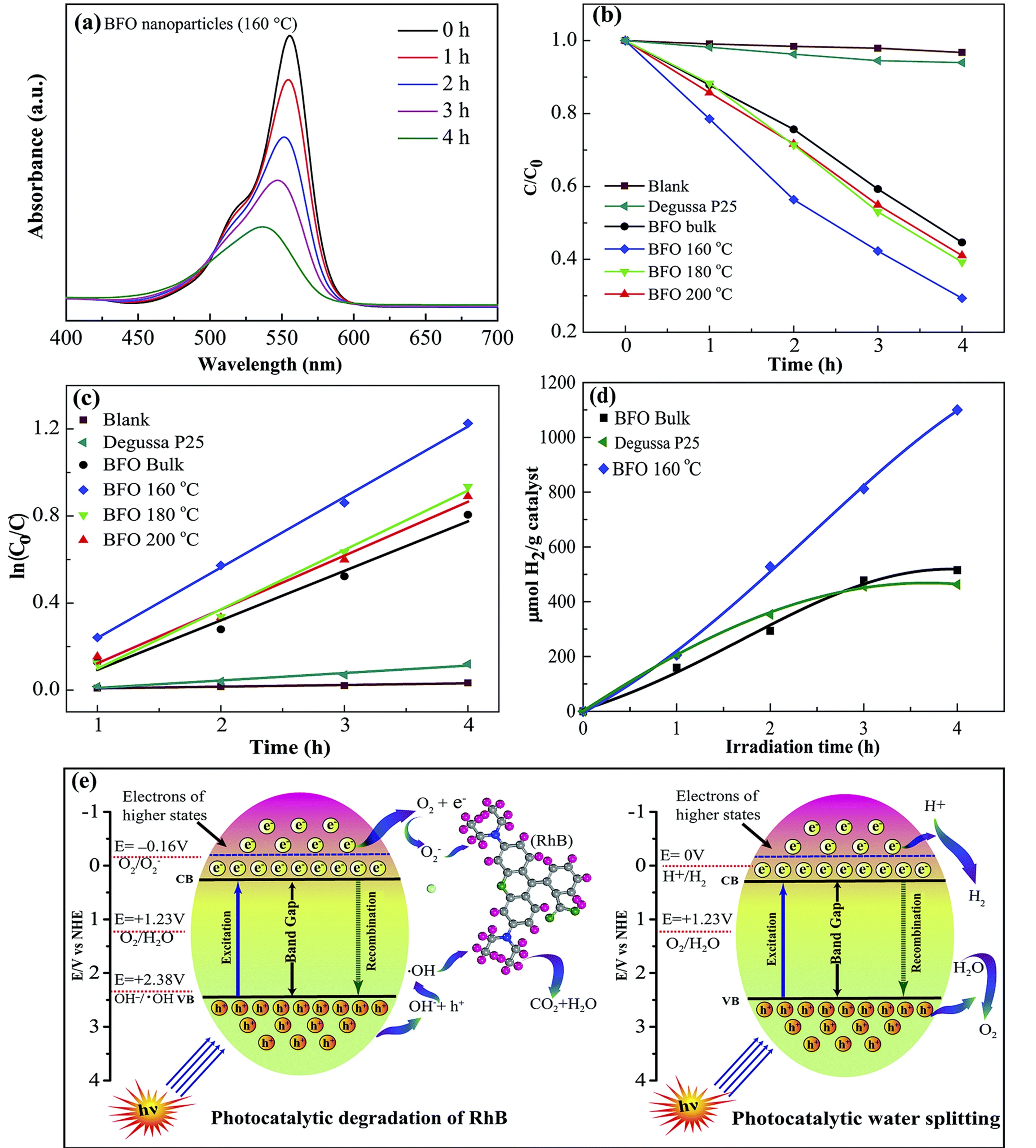}
 \caption{(a) Absorption spectra of rhodamine B (RhB) under visible-light irradiation in the presence of BFO nanoparticles prepared at 160 $^\circ$C hydrothermal reaction temperature. (b) Photocatalytic degradation of RhB as a function of the visible light irradiation time for blank sample, Degussa P25 titania, BFO bulk material and hydrothermally prepared nanoparticles at reaction temperatures 200 $^\circ$C, 180 $^\circ$C and 160 $^\circ$C. (c) Pseudo-first order kinetics fitting data for the photodegradation of RhB. (d) Volume of H\textsubscript{2} evolution as a function of irradiation time during the photocatalytic splitting of water. (e) Schematic illustration of RhB degradation and water splitting mechanism using BFO nanoparticles as catalysts under visible light irradiation.}
 
 \label{fig:fig8}
\end{figure*}

The photocatalytic performance  of the fabricated materials was studied extensively through degradation of the typical organic contaminant Rhodamine-B (RhB) under visible light illumination \cite{ref1405} at ${\lambda}$ $\geq$ 420 nm. Fig. \ref{fig:fig8} (a) shows how intensity of the absorbance peak of RhB for BFO 160 nanoparticles typically varies under visible-light irradiation. The absorbance peak corresponding to RhB decreases gradually with increasing time, implying that RhB has been decomposed by BFO. Due to the well known fact that RhB dye shows a general resistance to decomposition, the implication of the results is that the photocatalytic activity of BFO nanoparticles is quite efficient and hence indicates the promise as visible light photocatalysts for BFO nanoparticles.

\tab For evaluating the efficiency of degradation of RhB, the maximum intensity ratio C/C$_0$ is plotted in Fig. \ref{fig:fig8} (b), in which C$_0$ and C are the maximum initial intensity (0 h) and maximum intensity at a specific time (1-4 h) respectively for the absorption spectra of RhB under stimulated sunlight irradiation. A blank test is evaluated for RhB (Fig. \ref{fig:fig8} (b)) and it exhibits a negligible degradation efficiency which clearly shows a nominal self-degradation potential of RhB. For example, after 4 hours of irradiation without BFO materials, RhB was degraded by less than 3\%, whereas the bulk BFO photocatalyst could decompose 56\% of RhB after 4 h irradiation. Notably, when hydrothermally prepared nanoparticles were used as photocatalysts, the degradation efficiency was increased to 59 \% and 61 \% for nanoparticles prepared at 200 $^\circ$C and 180 $^\circ$C hydrothermal reaction temperatures. Interestingly, RhB photodegradation attains a significant improvement in efficiency to 79 \% for BFO 160 nanoparticles which were prepared at 160 $^\circ$C reaction temperature.

\tab In a similar investigation BiFeO$_3$ particles with the size of 100-150 nm were synthesized at the temperature as low as 120 $^\circ$C via a urea-assisted hydrothermal process \cite{ref9001}. Their photocatalytic activity was evaluated by the degradation of RhB in aqueous solution under visible-light irradiation using a 500 W Xe lamp with a cutoff filter (${\lambda}$ $\geq$ 420 nm). The condition for photocatalytic experiment was same as that of our experiment. Moreover, the optical band gap of the particles was 2.1 eV which is also consistent with that of the nanoparticles synthesized in the present investigation. However, in Ref. [\cite{ref9001}] it was observed that only 40\% of RhB was decomposed after 4 h visible-light irradiation. Though the band gap of the BFO particles synthesized both in Ref. [\cite{ref9001}] and in our investigation was same, however, the photoctalytic dye degradation is much higher while BFO 160 nanoparticles are used as photocatalysts. In another recent investigation \cite{ref9002}, the photocatalytic activity of BFO nanoparticles with a particle size of 40 nm was evaluated by photocatalytic decomposition of RhB in aqueous solution under visible light irradiation. It is observed that the pure BFO photocatalyst could decompose 22.3\% of RhB after 270 min visible light irradiation.

\tab The photodegradation of BFO 160 nanoparticles is much higher compared to that of other materials in particular commercially available Degussa P25 titania nanoparticles and BFO bulk materials. For comparison, the photodegradation efficiency of BFO 140 and BFO 120 materials was also evaluted under the same experimental condition. The degradation efficiency was found to 53 \% and 42 \% for using BFO 140 and BFO 120 particles, respectively. Our investigation clearly demonstrated that the photocatalytic activity of BFO 160 nanoparticles for photocatalytic decomposition of RhB in aqueous solution under visible light irradiation is higher compared than that of other materials under investigation. The enhanced photodegradation efficiency of BFO 160 nanoparticles might have resulted from the reduced particle size with excellent crystallinity. Looking at the SEM image, Fig. \ref{fig:fig2} (e-f), significantly greater surface area of the BFO 160 nanoparticles can be attributed to the higher photodegradation efficiency. It should also be noted that compared to commercially available photocatalyst TiO$_2$, that responds only to UV irradiation, BFO nanoparticles are more advantageous as they make use of the visible portion of the spectrum.

\tab For quantitative investigation of the rate of degradation, the experimental data were fitted adapting a first-order model as elucidated by the following equation, ln(C$_0$/C) = kt \cite{ref801}, where C$_0$ and C denote the respective concentrations of RhB at distinct irradiation times of t$_0$ and t, k is the first-order rate constant and can be regarded as a basic kinetic parameter for a number of different photocatalysts. For the photodegradation of RhB, the pseudo-first order kinetics fitting data has been shown in Fig. \ref{fig:fig8} (c). The corresponding reaction rate constant (k) values were calculated to be 7.66 x 10$^{-3}$, 2.27 x 10$^{-1}$, 2.47 x 10$^{-1}$, 2.72 x 10$^{-1}$, and 3.23 x 10$^{-1}$ min$^{-1}$ for the blank sample, bulk BFO, hydrothermally prepared nanoparticles at reaction temperatures 200 $^\circ$C, 180 $^\circ$C and 160 $^\circ$C, respectively. Among them, the BFO 160 nanoparticles exhibited the highest k value, which is 1.42 times higher than that of bulk BFO. Notably, BFO 160 nanoparticles were subject to recyclability test under the same test conditions. The test result indicates excellent stability of these nanoparticles after 4 cycles.

\begin{table}[t]
	\centering
\caption{Calculated band gap energy (E$_g$), rate constant (K), and degradation efficiency of BFO bulk and hydrothermally prepared nanoparticles.}\label{Table:xrd1}
	\begin{tabular}{|l|l|l|l|}
		\hline
		Sample & E$_g$ & Rate constant&Degradation\\  
		& in eV&(K) x 10 $^{-1}$& (\%) after 4 h \\  \hline
		BFO Bulk&2.1&2.27&56\\ \hline
		BFO (200 $^\circ$C)&2.05&2.47&59\\ \hline	
		BFO (180 $^\circ$C)&2.11&2.72&61\\ \hline
		BFO (160 $^\circ$C)&2.1&3.23&71\\ \hline
	\end{tabular}
\end{table}

\subsection{Photocatalytic Hydrogen Production}
Now, we have examined the potential of BFO materials for hydrogen generation via water splitting under visible light irradiation \cite{ref11001}. No hydrogen evolution could be detected when a blank experiment without any photocatalyst was performed in dark conditions. Fig. \ref{fig:fig8} (d) shows the hydrogen evolution rate of BFO bulk material as well as  hydrothermally prepared BFO 160 nanoparticles in mL H$_2$/g catalyst plotted against visible light irradiation time in hours (h). For comparison, we have inserted in Fig. \ref{fig:fig8} (d) the hydrogen production rate of commercially available Degussa P25 titania nanoparticles. The outcome of our investigation demonstrates that  BFO 160 nanoparticles prepared hydrothermally at reaction temperatures 160 $^\circ$C generate nearly two times hydrogen compared to that of bulk BFO material after 4 h of illumination. Compared to commercially available TiO$_2$, the production rate of BFO 160 nanoparticles is more than double after 4 h of illumination. In a separate investigation \cite{ref9}, BFO nanoparticles as photocatalyst were also used for solar hydrogen generation from water and the production rate after 4 h illumination is equivalent with that of our investigation. But here the noteworthy point is that, in Ref. [\cite{ref9}], the nanoparticles of size 50-60 nm were produced at comparatively high temperature sol-gel process. Particularly the nanoparticles used for hydrogen production in Ref. [\cite{ref9}] was subject to annealing at 600 $^\circ$C for 2 hours in N$_2$ atmosphere. Whereas in the present investigation a facile, low cost hydrothermal process has been used for fabrication of nanoparticles and the reaction temperature was only 160 $^\circ$C which ensures the less energy consumption.

\subsection{Photocatalytic Mechanism}

The photocatalytic performance under visible light irradiation depends on a number of parameters cooperating with each other ultimately ameliorating the photocatalytic performance. Fig. \ref{fig:fig8} (e) schematically illustrates degradation of RhB and water splitting mechanism using BFO nanoparticles under visible light irradiation \cite{ref11002}.

\tab Photocatalysis is an electrochemical process that involves the transfer of the photogenerated electrons and holes between a semiconductor and an electrolyte. The semiconductor in this investigation is BFO, and the electrolyte is the RhB solution. When the BFO nanoparticles are dispersed in the RhB solution under visible light irradiation, the electrons in their valence band are excited to the conduction band by absorbing the photons of the visible light. This transition creates holes in the valence band of BFO. On the other hand, an electrolyte in equilibrium has one or more redox couples with specific redox potentials. If the potential of the photogenerated electrons in a semiconductor is lower than any of these redox couples, the electrons will flow into the electrolyte to perform reduction of that couple. Similarly, if the potential of the photogenerated holes of the semiconductor is higher than a redox couple, the holes will flow into the electrolyte to oxidize it. The reduced and oxidized species can initiate further reactions that are responsible for degrading the electrolyte. Besides determining the potentials of the conduction band minima (CBM) and valence band maxima (VBM) of BFO, we need to investigate the redox potentials of the redox couples that dictate the degradation of RhB to learn about the photocatalytic mechanism behind the degradation of RhB. The potentials of CBM and VBM of BFO have been determined to be 0.3 V and 2.4 V respectively using the absolute electronegativity theory by Nethercot and Butl and et al. \cite{refNeth,refBut}.

\tab The degradation of RhB dye depends primarily on two critical redox reactions. First, the photogenerated electron can react with the surface adsorbed O$_2$ (redox potential: -0.16 V vs. NHE) to form O$_2^{-}$, which will further react with RhB to cause degradation \cite{refXXX,refXXXXX}. Second, the photogenerated holes can react with the OH$^{-}$ ionized from the water molecules to produce $\cdot OH$ (redox potential: 2.38 V vs. NHE). This $\cdot OH$ can further oxidize the RhB molecules. Therefore, a photocatalyst needs to possess a CBM \textless -0.16 V to drive the first reaction and VBM \textgreater 2.38 V to drive the second one efficiently. Therefore, only the photogenerated holes are capable of performing the second reaction. However, when photons with energy higher than 2.56 eV are incident on BFO nanoparticles, the electrons in their valence band can be excited to the higher states of the conduction band where their potential can be more negative than -0.16 V. These photogenerated electrons will be able to perform the first reaction. Since BFO can absorb visible light photons with energies higher than 2.56 eV, both these reactions can contribute to the degradation of RhB \cite{ref11109}. In addition, RhB self-photoexcitation may also play a role in assisting the photodegradation process. The RhB molecules can absorb the visible range photons to reach an excited state of a potential of -1.09 V \cite{refNANO}. Since the potential of electrons in this state is more negative than the CBM of BFO nanoparticles, the electrons of the photoexcited state will be transferred to the conduction band of BFO. However, these electrons cannot participate in the photocatalytic degradation process as they cannot reduce O$_2$ to O$_2^{-}$. Consequently, these electrons will be accumulated in the conduction band of BFO and the transfer of electrons from the photoexcited state of RhB will stop when both states reach the same potential. Hence, the self-photoexcitation of RhB may not have a significant effect on the photocatalytic degradation of RhB.

\tab It is well known that the size of the particles and surface area significantly effect the enhancement of the photocatalytic degradation. In our investigation, the degradation of RhB with BFO 160 nanoparticles is considerably greater than that of bulk BFO owing to their smaller size and higher surface area. The band gap is another important parameter as the low band gap of photocatalysts might increase the photocatalytic activities \cite{ref2004}. We observed that the band gap is almost same (2.1 eV) for nanoparticles prepared at all hydrothermal reaction temperatures as well as the bulk BFO. While sunlight hits the photocatalyst, photons possessing greater energy than the optical band gap of the photocatalyst cause stronger light absorption. The estimated band gap 2.1 eV for both BFO bulk and nanoparticles implied that light with $\lambda$ less than 590 nm could be absorbed by e\textsuperscript{--} to jump from valence band (VB) to conduction band (CB) which covers a broad region of the solar energy spectrum.  With all the materials under scrutiny having almost equal band gap, the other factors prove to be more important for achieving higher photocatalytic efficiency of BFO 160 nanoparticles rather than band gap itself.

\tab It was mentioned already in introduction section that, BFO is an well known multiferroic materials.  It is evident that the ferroelectric property of a material enhances its ability to decolorize Rhodamine-B, a typical dye molecule, while irradiated by solar light \cite{ref1001}. Ferroelectric materials have a spontaneous polarization that is caused by the non-centrosymmetry of the crystal structure. As nanoparticles are more strained than their bulk counterparts, their crystal structures deviate from the bulk structure and give rise to even greater non-centrosymmetry for BFO. This essentially increases the magnitude of spontaneous polarization of BFO nanoparticles, especially for BFO 160 with smallest particle sizes as evident from our experimental values. The BFO 160 nanoparticles, assumed to be creating electrical field in the surrounding medium, improves redox reactions along with the adsorption of dye molecules, especially polar molecules such as water on the photocatalyst surface. This adsorption helps the redox reactions to commence easily as the carrier transport is now relatively easier. Furthermore, it is anticipated that ferroelectrics can create channels to transfer the charge carriers to the surface of photocatalyst and significantly reduce the recombination probabilities as the photogenerated electron-hole pairs are driven to opposite directions by the electric field they possess.

\section{Conclusions}

We have produced single-phase, well crystalline BiFeO$_3$ nanoparticles with favorable morphology using a facile, low cost hydrothermal process at reaction temperature as low as 160 $^\circ$C. The experimentally observed band gap (2.1 eV) was almost equal for both bulk and all nanoparticles prepared at different reaction temperatures and was well consistent with value obtained from the first principles calculation. However, the BFO 160 nanoparticles prepared at temperature 160 $^\circ$C demonstrated much higher photocatalytic activity compared to the photocatalytic performance showed by bulk BFO and other nanoparticles. The enhanced photocatalytic activities of BFO 160 nanoparticles may be associated with their crystallinity, phase purity, excellent morphology, higher surface to volume ratio, and the efficient separation and migration of photogenerated charge carriers. As efficient photocatalyst, the synthesized BFO 160 nanoparticles generated more than two times of solar hydrogen via water splitting than that of bulk BiFeO$_3$ as well as commercially available Degussa P25 titania. The outcome of this investigation also demonstrated an improved multiferroic properties of technologically important BiFeO$_3$ nanoparticles prepared at temperature as low as 160 $^\circ$C which is cost effective with less energy consumption. BiFeO$_3$ nano-structured particles prepared through this investigation have shown a great promise in enhancing the production of solar H$_2$, a carbon free fuel, using two important renewable sources: water and solar energy.

\section{Acknowledgements}
This work was financially supported by Ministry of Education, Government of Bangladesh (Grant No. PS 14267) and the Infrastructure Development Company Limited (IDCOL), Dhaka, Bangladesh.

\section{Appendix A. Supplementary data} 
 Supplementary data to this article can be found online.

\section{Data availability}
The raw and processed data required to reproduce these findings cannot be shared at this time due to technical or time limitations.


\bibliography{PCCP} 
\bibliographystyle{rsc} 


\end{document}